\documentclass[a4paper,12pt]{article}
\usepackage[pctex32]{graphics}

\begin{document}

\topmargin 0pt \oddsidemargin 0mm
\newcommand{\beq}{\begin{equation}}
\newcommand{\eeq}{\end{equation}}
\newcommand{\beqa}{\begin{eqnarray}}
\newcommand{\eeqa}{\end{eqnarray}}
\newcommand{\fr}{\frac}

\begin{titlepage}
\begin{flushright}
{\tt gr-qc/0609031}
\end{flushright}

\vspace{5mm}
\begin{center}
{\Large \bf Black hole thermodynamics with generalized uncertainty
principle } \vspace{12mm}

{\large   Yun Soo Myung$^{\rm a,}$\footnote{e-mail
 address: ysmyung@inje.ac.kr},
 Yong-Wan Kim $^{\rm b,}$\footnote{e-mail
 address: ywkim@pcu.ac.kr},
and Young-Jai Park$^{\rm c,}$\footnote{e-mail
 address: yjpark@sogang.ac.kr}}
 \\
\vspace{10mm} {\em $^{\rm a}$Institute of Mathematical Science and
School of Computer Aided Science, \\Inje University, Gimhae
621-749, Korea \\} {\em $^{\rm b}$ National Creative Research
Initiative Center for Controlling Optical Chaos, \\Pai-Chai
University, Daejeon 302-735, Korea \\} {\em $^{\rm c}$Department
of Physics  and Center for Quantum Spacetime,\\ Sogang University,
Seoul 121-742, Korea}

\end{center}

\vspace{5mm} \centerline{{\bf{Abstract}}}
 \vspace{5mm}
We apply  the generalized uncertainty principle to the
thermodynamics of a small black hole. Here we have a black hole
system with the UV cutoff.  It is shown that the minimal length
induced by the GUP interrupts the Gross-Perry-Yaffe phase
transition for a small black hole. In order to see whether the
black hole remnant takes place a transition to a large black hole,
we introduce a black hole in a cavity (IR system). However, we
fail to show the phase transition of the remnant to the large
black hole. \vspace{3mm}


\end{titlepage}

\newpage
\renewcommand{\thefootnote}{\arabic{footnote}}
\setcounter{footnote}{0} \setcounter{page}{2}

\section{Introduction}
Hawking's semiclassical analysis of  the black hole  radiation
suggests that most  information about initial states is shielded
behind the event horizon and will not back to the asymptotic
region far from the evaporating black hole~\cite{HAW1}. This means
that the unitarity is violated by an evaporating black hole.
However, this conclusion has been debated  by many authors for
three decades~\cite{THOO,SUS,PAG}. It is closely related  to the
information loss paradox which states the question of whether the
formation and subsequent evaporation
 of a black hole is unitary. One of the most urgent problems in the black
hole physics is  the lack of resolution of  the unitarity issue.
Moreover, a complete description of the final stage of the black
hole evaporation is important but is still quite unknown. To reach
the solution to these problems, we need to use quantum gravity.

 Although two
leading candidates for quantum gravity are the string theory and
the loop quantum gravity, we need to introduce an alternative
approach that provides a manageable form of the quantum gravity
effect. It was proposed that the holographic principle could serve
such a purpose because it includes the effect of the quantum
mechanics and gravity~\cite{Ho,Su}. Nowadays we are interested in
the generalized uncertainty principle (GUP) and its
consequences~\cite{nc,GUP,ACS} since the Heisenberg principle is
not expected to be  satisfied when quantum gravitational effects
become important. Even though the GUP has its origins in the
string theory~\cite{mo} and it is similar to
noncommutativity~\cite{nc}, the principle provides the minimal
length scale and thus modifies the thermodynamics of the black
hole at the Planck scale significantly. One  important
modification is how the GUP prevents
 total Hawking evaporation of the small black hole similar to
the way the Heisenberg principle prevents the hydrogen atom from
total collapse. Hence, there remains a remnant of  a black hole.
The authors~\cite{ACS} insisted that  the remnant does not
necessarily have a  black hole horizon. Thus they have set the
entropy  to  zero at the Planck scale. However, this approach is
not a complete work of  the GUP effect on  black hole
thermodynamics. We call the thermodynamic picture modified by the
GUP the UV thermodynamics of a black hole.

On the other hand,  there exists the Gross-Perry-Yaffe (GPY) phase
transition from thermal radiation to a black hole~\cite{GPY}.
However, a black hole nucleated at the Hawking temperature
$T_{H}=1/8\pi M$ is in an unstable equilibrium with a heat
reservoir of the same temperature $T=T_{H}$ in asymptotically flat
spacetime. Its fate, under small fluctuations, will be either to
decay to hot flat space by radiating or to grow without limit by
absorbing radiation in the heat reservoir.  There is a way to
achieve a stable large black hole in equilibrium with a heat
reservoir. A black hole could be rendered thermodynamically stable
by introducing a cavity~\cite{York,WY,Brown,AD}. An important
point is to know how the system of a black hole with positive heat
capacity could be realized  through the phase transition. This
corresponds to the IR thermodynamics of the black hole in a
cavity. We call  the phase transition between thermal radiation
and the large black hole in a cavity as the well-defined
Gross-Perry-Yaffe (WGPY) transition\footnote{Actually, this
transition corresponds to the Hawking-Page transition in the anti
de Sitter spacetime~\cite{HP}.}. The WGPY transition could be
checked by observing the heat capacity and free energy~\cite{SH}.

Our work is based on the first law of $dE=TdS$ rather than $dM=TdS$.
We investigate  the UV thermodynamics by introducing the thermal
energy $E$ instead of the Arnowitt-Deser-Misner (ADM) mass $M$. We
find that the black hole remnant is a thermodynamically stable
object because of its positive heat capacity. Furthermore, we
introduce the geometric means of UV and IR thermal quantities to
study whether the phase transition to a stable large black hole is
possible.

 The organization of this work is as follows.
We first study the UV thermodynamics for a  modified black hole
by the GUP in section II.
  Section III is devoted to reviewing
 the WGPY phase transition of the Schwarzschild black hole in a
 cavity,
which corresponds to the IR thermodynamics for the large black
hole.
 Finally, we discuss  and  summarize our results  of thermodynamic properties in section IV.
\section{GUP and UV thermodynamic system}
The GUP possesses a minimal length scale which leads to a finite
resolution of the spacetime. Thus, we could use the GUP to modify
thermodynamic quantities of the black hole. Let us start with the
GUP \beq \label{1eq2} \Delta x \ge \frac{\hbar}{\Delta p}+l_p^2
\frac{\Delta p}{\hbar} \eeq with $l_p=\sqrt{G \hbar/c^3}$ the
Planck length. The Planck mass is given by $M_p=\sqrt{\hbar c/G}$.
The above implies a lower bound on the length scale \beq
\label{2eq2} \Delta x \ge 2 l_p, \eeq
 which means that the Planck length plays
an  important role as a fundamental scale. On the other hand,
 the GUP may be used to derive the modified black hole temperature. A
simple calculation provides the modified temperature for radiated
photons. The momentum uncertainty for radiated photons can be
found to be

\beq \label{3eq2}
 \frac{\Delta p}{\hbar}=\frac{\Delta x}{2 l_p^2}
  \Big[1\pm \sqrt{1-\frac{4l_p^2}{(\Delta
 x)^2}}\Big].
 \eeq
For simplicity we use the Planck units of $c=\hbar=G=k_B=1$  which
imply that $l_p=M_p=1$. Considering the GUP effect near horizon and
$\Delta x=r_{S}=2M$, the energy (temperature) of radiated photons
$\Delta p$ can be identified with the UV temperature of a black hole
up to a factor of $2\pi$ \beq \label{4eq2}
 T_{UV}=\frac{M}{4\pi}
  \Big[1- \sqrt{1-\frac{1}{M^2}}\Big].
 \eeq
Here we choose a negative sign to recover the Hawking temperature
of $T_{H}$ in the limit of   large  $M$. We are now in a position
to introduce the UV thermal energy by analogy with the IR
thermodynamics~\cite{York,WY,Brown} \beq \label{5eq2}
E_{UV}=2M^3\Big(1-\sqrt{1-\frac{1}{M^2}}\Big). \eeq In the case of
$M \gg 1$, this energy approximates to $E_{UV} \simeq
M\Big(1+1/4M^2)$. It takes the Schwarzschild mass of $E_{S}= M$ in
the limit of large $M$. Rewriting $M$ in terms of $E_{UV}$, we
have $M\simeq E_{UV}(1-\sqrt{1-1/E_{UV}^2})/2$, which shows that
the ADM mass  consists of the thermal energy and gravitational
self-energy associated with the GUP\footnote{We note that the
difference between  mass and thermodynamic energy was first
clarified in \cite{BY}.}.

The entropy correction is obtained by integrating the first law of
$dE_{UV}=T_{UV}dS_{UV}$.
 The modified entropy
thus is given by \beq \label{6eq2}
 S_{UV}=4 \pi \Big[M^2\Big(2- \sqrt{1-\frac{1}{M^2}}\Big)-
 \ln\Big(M+\sqrt{M^2-1}\Big)\Big],
 \eeq
which approaches  the  Bekenstein-Hawking entropy $S_{BH}=4 \pi
M^2$ in the limit of large $M$. The modified heat capacity takes
the form \beq \label{7eq2}
 C_{UV}=\frac{dE_{UV}}{dT_{UV}}=8 \pi \Big(M^2-2M
 \sqrt{M^2-1}\Big),
 \eeq
 which  leads to the heat capacity for
the Schwarzschild black hole $C_{S}=-8\pi M^2$  in the limit of
large $M$.
 Finally, the on-shell free
energy is defined by \beq \label{8eq2}
 F^{on}_{UV}=E_{UV}-T_{UV}S_{UV}.
 \eeq

\begin{figure}[t!]
   \centering
   \includegraphics{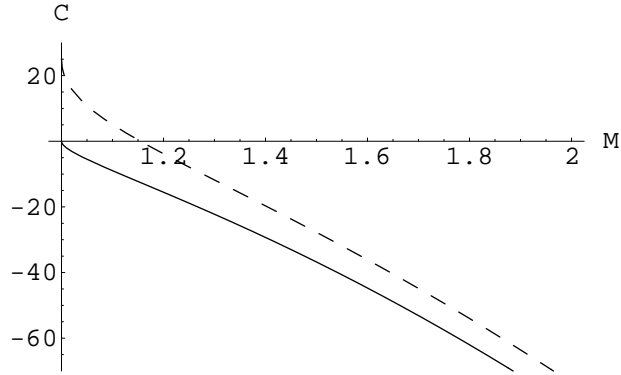}
\caption{The dashed line:  heat capacity $C_{UV}$, and  the solid
line: heat capacity $\tilde{C}_{UV}$ as a function of the black hole
mass $M$.   } \label{fig1}
\end{figure}

\begin{figure}[t!]
   \centering
   \includegraphics{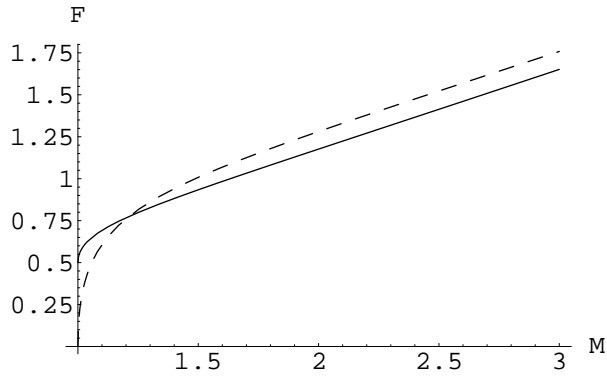}
   \caption{Plot of the on-shell free energies as a function of $M$.
    The dashed line denotes
    $F^{on}_{UV}$, and the solid line represents $\tilde{F}^{on}_{UV}$. } \label{fig2}
\end{figure}
 At this stage, it is appropriate  to comment on the difference between the ADM mass $M$ and thermal energy $E_{UV}$.
 If one uses  $M$ instead of  $E_{UV}$, the heat capacity\footnote{Here
 we distinguish the case with mass $M$ from the case with the thermal energy $E$
  by introducing the notation of $\tilde{{\cal O}}$: $(C_{UV},S_{UV},F_{UV})$ and
  $(\tilde{C}_{UV},\tilde{S}_{UV},\tilde{F}_{UV})$.} is given by
  \beq \label{9eq2}
 \tilde{C}_{UV}=\frac{dM}{dT_{UV}}=4\pi
 \frac{\sqrt{1-\frac{1}{M^2}}}{\sqrt{1-\frac{1}{M^2}}-1}.
\eeq Also  the free energy takes the form
 \beq \label{10eq2}
 \tilde{F}^{on}_{UV}=M-T_{UV}\tilde{S}_{UV}
 \eeq
with the unnormalized entropy derived from the integration of
$dM=T_{UV}d\tilde{S}_{UV}$~\cite{ACS}\beq \label{11eq2}
 \tilde{S}_{UV}=2 \pi \Big[M^2\Big(1+ \sqrt{1-\frac{1}{M^2}}\Big)-
 \ln\Big(M+\sqrt{M^2-1}\Big)\Big].
 \eeq
In the limit of  large $M$, one  recovers  the  free energy
$F^{on}_{S}=M-T_{H}S_{BH}=M/2$ for the Schwarzschild black hole
from both $F^{on}_{UV}$ and $\tilde{F}^{on}_{UV}$. As is shown in
Fig. 1, we find that $\tilde{C}_{UV}(M=1)=0$, while $C_{UV}(M=1)=8
\pi>0$. We observe that a thermodynamically unstable region
($C_{UV}\le 0$) appears for $M \ge 1.15$. This means that the
remnant of black hole  is thermodynamically stable near the Planck
scale if one takes  a proper definition of the energy into
account. Also we find from Fig. 2 that the GPY phase transition
from thermal radiation ($M=0$)
  to a small black hole  is impossible because there exists a forbidden region
  of $0\le M<M$. We suggest  that the phase transition from the remnants ($M=1$) to a
large black hole is possible to occur. However, this fact is
unclear until one studies the
 a black hole in a
cavity.

\section{IR system: Schwarzschild black hole in a cavity}

A black hole of mass $M$ can be rendered thermodynamically stable
by confining it within an isothermal cavity. We assume that a
black hole is located at the center of the cavity.  Here we fix
the temperature $T$ on its isothermal boundary of radius $R$. In
an equilibrium configuration, the Hawking temperature $T_{IR}$
measured on the boundary must be equal to the boundary temperature
$T$~\cite{York} \beq T_{IR}\equiv \frac{1}{8\pi
M}\frac{1}{\sqrt{1-\frac{2M}{R}}}=T. \label{1eq3}
 \eeq
This means that, according to  the Tolman law,  a local observer
at rest will measure a local temperature $T$ which scales as
$1/\sqrt{-g_{00}}$ for any self-gravitating system in thermal
equilibrium with a  heat reservoir. The cavity is regarded as the
heat reservoir. At $M=M_{IR0}=R/3$, $T_{IR}$ has the minimum
temperature
 \beq
 \label{2eq3}
 T_{IR0}=\frac{\sqrt{27}}{8\pi R},
 \eeq
 which corresponds to the nucleation temperature of a black hole.
 The nucleation is a purely quantum-gravitational phenomenon.
 The  equation (\ref{1eq3}) allows two
solutions for a given $T$: an unstable small black hole  with mass
$M_u$ and a stable large black hole with  mass $M_s$. For
$T<T_{IR0}$, no real value of $M$ can solve Eq.(\ref{1eq3}),
making it hard for any black hole  to exist in the cavity. The
solutions in the left-handed side of $M_{IR0}$ correspond to
unstable black holes with mass $M_u\simeq (1/8 \pi T)[1+1/8\pi
RT]$, while those in the right-handed side are stable black holes
with mass $M_s\simeq (R/2)[1-1/(4\pi RT)^2]$~\cite{York}. $M_s$
arises from the effect of a cavity. Here we  have a sequence of
$M_u<M_{IR0}=33<M_s$ for $R=100$.

 The thermal energy is derived
from the first law of $dE_{IR}=T_{IR}dS_{IR}$ by assuming that the
black hole entropy is not changed. The thermal energy and entropy
for the black hole embedded in a cavity take the form of
$E_{IR}=R\Big(1-\sqrt{1-2M/R}\Big)$ and $S_{IR}=4 \pi M^2$,
respectively. Solving $E_{IR}$ for $M$ leads to
$M=E_{IR}-E_{IR}^2/2R$, which states that the ADM mass consists of
the thermal energy and  the gravitational self-energy due to the
presence of a cavity. The heat capacity is defined as
$C_{IR}\equiv (\partial E_{IR}/\partial T_{IR})_A$ at the constant
area $A=4 \pi R^2$ of the  boundary. The heat capacity and free
energy $F^{on}_{IR}\equiv E_{IR}-T_{IR}S_{IR}$ are given as \beq
 \label{3eq3}
C_{IR}=-\frac{8 \pi
M^2(1-2M/R)}{(1-3M/R)},~~F^{on}_{IR}=R\Big(1-\sqrt{1-2M/R}\Big)-\frac{M}{2\sqrt{1-2M/R}}.
 \eeq
Actually, $F^{on}_{IR}=0$ leads to the phase transition at
$M=M_{IR1}=4R/9$. Substituting it into Eq.(\ref{1eq3}), the phase
transition temperature can be obtained as \beq \label{4eq3}
T_{IR1}=\frac{27}{32 \pi R}. \eeq

As is shown in Eq.(\ref{3eq3}), $C_{IR}$ has a singular point  at
$M=M_{IR0}$. This  suggests that the  phase transition  is the
first-order at $T=T_{IR0}$. However, this behavior of heat
capacity  does not indicate a phase transition in the canonical
ensemble. The heat capacity determines thermal stability of the
system. The system is thermally unstable for $M<M_{IR0}$, while it
is stable for $M>M_{IR0}$. On the other hand, the IR-free energy
$F^{on}_{IR}$ does not have a pole at $M=M_{IR0}$. Instead, it has
the maximum value at this point and  becomes zero at $M=M_{IR1}$.
\begin{figure}[t!]
\centering
   \includegraphics{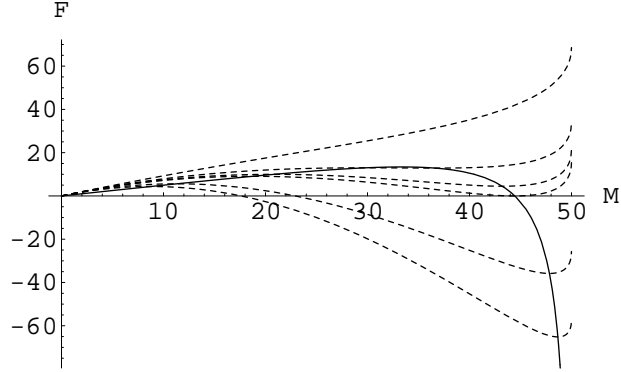}
   \caption{The solid line represents the IR-free energy $F^{on}_{IR}$, while
the dashed lines denote the off-shell free energy $F^{off}_{IR}$.
From the top down, we have the off-shell free energy graphs for
$T=0.001, T_{IR0}(=0.0021), 0.0025, T_{IR1}(=0.0026), 0.004,
0.005$.} \label{fig3}
\end{figure}

In order to study the phase transition explicitly, let us
introduce the off-shell free energy, which  plays a role of
effective potential in the canonical ensemble, as follows
~\cite{York,SH,Myung1,Myung2}: \beq \label{5eq3}
F^{off}_{IR}=E_{IR}-TS_{IR}=R\Big(1-\sqrt{1-2M/R}\Big)-4 \pi M^2
T. \eeq  With $T=T_{IR0}$, an extremum appears at
$M=M_{IR0}(=M_u=M_s)$. This  could be checked by noticing an
inflection point in Fig. 3. The nucleation of a stable black hole
with mass $M_s$ occurs at $T=T_{IR0}$. For $T>T_{IR0}$, there are
two extrema, the unstable small black hole with mass $M_u$ and the
stable large  black hole with mass $M_s$.  We note that for
$T_{IR0}<T<T_{IR1}$, $F^{off}_{IR}$ has a saddle point (not
maximum) at $M=M_u$. This unstable solution is important as the
mediator of the  phase transition from thermal radiation to a
stable black hole. At $T=T_{IR1}$, there is a transition between
thermal radiation and a  large black hole. That is, for
$T<T_{IR1}$ thermal radiation dominates, while for $T>T_{IR1}$ a
black hole dominates. This is a picture of the WGPY phase
transition: $M=0 \to M=M_u \to M_s$. Furthermore, $F^{on}_{IR}$ is
a set of stationary points of $F^{off}_{IR}$. That is,
$F^{on}_{IR}$ can be obtained from the off-shell free energy
$F^{off}_{IR}$ through the dynamic operation: $\partial
F^{off}_{IR}/\partial M=0 \to T=T_{IR} \to
F^{off}_{IR}=F^{on}_{IR}$. This shows why we use the name of
``off-shell" clearly.

\section{Discussion and Summary}

\begin{figure}[t!]
   \centering
   \includegraphics{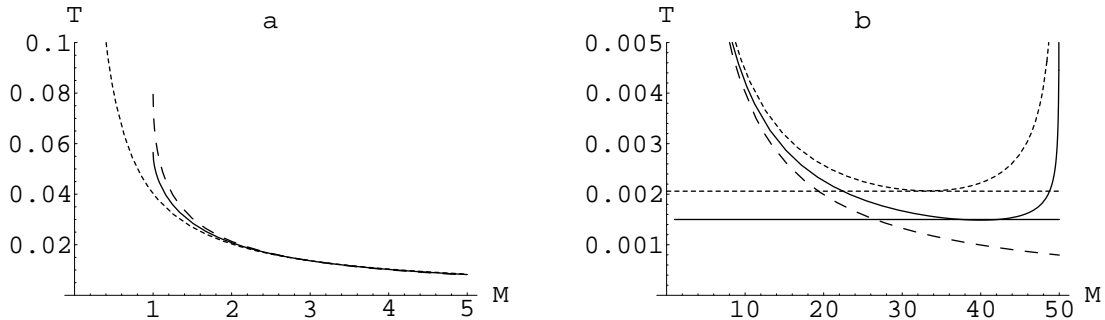}
\caption{ The graphs of temperature:  a) for $0\le M \le 5$, and
b) for $0\le M \le 50$.  Solid line:  Hawking temperature $T_{GM}$
of the geometric mean.
 Short-dashed line: temperature $T_{IR}$ of the IR
system. Long-dashed line: temperature $T_{UV}$ of the UV system
inspired by the GUP.} \label{fig4}
\end{figure}

The WGPY phase transition occurs between  hot flat space (thermal
radiation) and  a large  black hole. Its key feature is  the
transition from
 negative heat capacity to  positive heat capacity by the introduction of  an
isothermal cavity. However, there is still a drawback concerning
the divergent behavior of the black hole temperature, which
appears at the final stage of the black hole evaporation process
as a result of  Hawking radiation.

 On the other hand, if one introduces the GUP
into the black hole thermodynamics, the minimal length of the
Planck scale turns on. This modifies thermodynamic quantities of
the black hole near the Planck scale significantly. As a result,
 a black hole remnant  remains with the maximum temperature, which
 can be expressed as $C_{UV}=8\pi>0, F^{on}_{UV}=0, T_{UV}=0.056$
(maximum) at the Planck scale. Surprisingly, there exists an
inaccessible region of $0\le M<1$, that  forbids the GPY phase
transition from thermal radiation ($M=0$) to a small black hole. A
similar case is found in the topological AdS black
holes~\cite{Myung3}.

Moreover, one could expect to encounter another phase transition
from the remnant at $M=1$ to a
 large black hole.
 For this purpose,
one  may attempt to incorporate the UV system with the IR system.
One ansatz is to use their geometric mean of ${\cal O}_{UV}$ and
${\cal O}_{IR}$ defined  by ${\cal O}_{GM}=\sqrt{{\cal
O}_{UV}{\cal O}_{IR}}$~\cite{BV,Pad}.  First, let us consider the
temperature of the black hole system with UV and IR cutoffs. We
have three temperatures: $T_{UV}, T_{IR},
T_{GM}=\sqrt{T_{UV}T_{IR}}$.  We find immediately  from Fig. 4a
that the geometric mean does not works  near the Planck scale
well. At this point, we note  that when $M$ is small, $T_{GM}$
differs from both $T_{UV}$ and $T_{IR}$.  As is shown in Fig. 4b,
the minimum temperatures of $T_{IR}$ and $T_{GM}$ are given by
$T_{IR0}(M=33)=0.0021$ and $T_{GM0}(M=40)=0.0015$. Thus, applying
the geometric mean leads to the unwanted result that the minimum
point is shifted from $M_{IR0}=33$ into $M_{GM0}=40$. It turns out
that this temperature is not a physical temperature but   a
mathematical device. As a result, we have failed to show that this
transition occurs naturally. We attribute this failure to our
choice of the geometric means to describe the transition.

In summary, we have shown that the minimal length induced by the
GUP interrupts the GPY phase transition for small black hole and
the phase transition  of the remnant  to a large black hole.

\newpage
\section*{Acknowledgement}
This work was supported by the Science Research Center Program of
the Korea Science and Engineering Foundation through the Center
for Quantum Spacetime of Sogang University with grant number
R11-2005-021.  Y.-J. Park was also in part  supported by the Korea
Research Foundation Grant funded by the Korean Government
(KRF-2005-015-C00105).

\end{document}